\documentstyle[eqsecnum,aps,prl,epsfig,amssymb]{revtex}
\begin{document}
\draft

\title{Tunneling through a multigrain system:\\
deducing the sample topology from the nonlinear conductance
}

\author{Andrey V. Danilov$^a$, Dmitrii S. Golubev$^b$, and Sergey E. Kubatkin$^a$}

\address{MINA, Chalmers University of Technology,
S-41296, G\"oteborg, Sweden\\
$^a$also at Kapitza Institute, 113117 Moscow, Russia\\
$^b$also at Lebedev Physics Institute, 117924, Moscow, Russia}

\maketitle

{\setlength\arraycolsep{2pt}

\begin{abstract}
We study a current transport through a system of a few grains connected with tunneling links.
The exact solution is given for an arbitrarily connected double-grain system with a shared gate 
in the framework of the orthodox model.
The obtained result is generalized for multigrain systems with strongly different tunneling resistances.
We analyse the large-scale nonlinear conductance and demonstrate how the sample topology can be 
unambiguously deduced from the spectroscopy pattern (differential conductance versus gate-bias plot).
We present experimental data for a multigrain sample and
reconstruct the sample topology. A simple selection rule is formulated to distinguish samples with 
spectral patterns free from spurious disturbance caused by recharging of some grains  nearby.
As an example, we demonstrate experimental data with additional peaks in the spectroscopy pattern, 
which can not be attributed to coupling to additional grains. The described approach can be used 
to judge the sample topology when it is not guaranteed by fabrication and direct imaging is not possible. 

\end{abstract}

\newpage

\section{INTRODUCTION}
The electrical properties of nanostructures with ever-smaller size have been studied 
intensively during the past decade in a search for new quantum effects and novel types of 
electronic devices. The scientific interest has moved towards objects with sizes of a few 
nanometers. Such objects cannot be prepared by conventional electron beam lithography with 
a resolution being limited to $\sim 20$~nm.
The laboratory methods, developed thus far, make use of pre-fabricated nanoscale 
objects: semiconductor~\cite{Klein} or metallic~\cite{Persson} nanocrystals made by colloidal chemistry, 
organometallic carborane clusters~\cite{Soldatov}, and the fullerene molecules~\cite{Park}.
The granular films formed due to surface atom diffusion at room 
temperature~\cite{Ralph95,Ralph97,Gueron,Davidovich,OurAPL} or in a process of an avalanche 
reconstruction of a quench-condensed precursor layer~\cite{OurJLTP} were also used as a natural 
source of nanoparticles.
These prefabricated nanoparticles 
have been spread on 
a surface and studied with STM~\cite{Soldatov}; 
or placed into a gap between two macroscopic electrodes defined by electron beam litography~\cite{Klein} 
or made by a break-junction technique~\cite{Park}.

The final sample geometry, i.e. the number of nanoparticles in a gap and their arrangement 
is not under experimental control.
With some luck, one can get a sample where a single nanograin bridges two macroscopic electrodes.
Typical yields of 25-30\%
have been reported~\cite{Ralph97,Davidovich}.

Several methods to get a better defined sample geometry were suggested.
A hole in a silicon-nitride membrane was used to restrict a contact area~\cite{Ralph95,Ralph97,Gueron}.
A self-assembling of gold clusters on electrodes covered with monolayers of organic molecules was 
used by Ref.~\cite{Persson}. 
An alternative way is to modify a sample by {\it in situ} material deposition, 
as described in~\cite{OurAPL,OurJLTP}.

Nevertheless, not a single method ensures, just by the fabrication procedure itself, that there 
is no more than one nanoparticle in a gap. Moreover, even if the current mainly goes through a 
single grain, there may be some additional grains nearby which can potentially affect transport 
properties due to capacitance coupling.
Sometimes, the sample geometry can be imaged directly by STM or SEM, like in~\cite{Klein,Davidovich}, 
but sometimes not, as the samples are metastable and can only exist at cryogenic temperatures~\cite{OurJLTP}, 
or nanoobjects are simply too small, like the fullerene molecules~\cite{Park}.

The only universal way to judge the sample topology is to analyze current transport data.
Several criteria have been suggested in the literature, all based on the Coulomb blockade phenomena~\cite{Ortho}.
A nano-scale object, coupled to macroscopic source and drain electrodes, is a Single Electron 
Transistor - SET (although not every method allows to implement a third electrode - gate).
The exact periodicity of the conductance variation versus bias and gate voltages in a single grain 
SET is violated if more than one grain is involved in the current transport~\cite{Ruzin}.
So one can formulate an intuitive "selection rule" for single grain samples as "the ones 
showing Coulomb-staircase structure"~\cite{Ralph97,Persson,Gueron} or, more precisely, 
"the ones with a single set of equally spaced Coulomb-staircase peaks in the differential 
conductance"~\cite{Ralph95}.

Within the scope of the orthodox model for SET~\cite{Ortho}, the Coulomb staircases are the only 
peculiarities in the differential conductance. The real nanoscale samples, however, 
never follow this model exactly.
First, if a typical level spacing $\delta$ between the quantised energy levels in a nanoparticle is 
comparable with the charging energy $E_{\mathrm C}$, then the Coulomb diamonds have different 
sizes and the Coulomb staircases are no more periodic~\cite{Kouwenhoven}.
Second, if the temperature is below the inter-level spacing $\delta$, then every discrete energy level in a 
grain gives an additional peak in the differential conductance ${\mathrm d}I/{\mathrm d}V$~\cite{Ralph95}.
In fact, it was the discrete excitation spectrum in a nanoparticle, which was the primary subject of 
the research in references~\cite{Park,Ralph95,Ralph97,Gueron,Davidovich}.

Therefore, one should distinguish between the peculiarities coming from the discrete energy levels and a 
potential disturbance due to additional grains nearby~\cite{Zeeman}.
If all energy levels in the grain energy spectrum are equally shifted by the gate potential then 
any discrete energy state 
is revealed in a ${\mathrm d}I/{\mathrm d}V(V,V_g)$ plot 
($V$ is the gate and $V_g$ is the bias voltage) 
as a straight line parallel to the diamond edge~\cite{Ralph97}. This allows us to reformulate the selection 
rule for single grain samples as "the ones with all spectral lines in a ${\mathrm d}I/{\mathrm d}V(V,V_g)$ 
plot parallel to the diamond edges".

However, the gate voltage shifts all discrete levels the {\it same} way if only 
the external electrostatic field is effectively screened within a thin surface layer $\Delta \ll R$, 
where $R$ is characteristic sample radius. This is valid in the extreme case of a macroscopic metallic 
sample, but definitely does not take place for the opposite extreme case - a fullerene molecule.
Moreover, additional spectral features can be expected due to excitation of mechanical 
degrees of freedom~\cite{Park,Schwabe}, including the shuttle effects predicted in~\cite{Gorelik}, and 
they could have a distinct gate dependency.
To summarize, no single qualitative criterion, suggested thus far, provides a conclusive argument 
to prove a single grain sample topology.

In this paper we report the exact solution (within an orthodox model) for an arbitrary double 
grain system, and some mathematical results concerning the multigrain systems. Based on our 
analysis, we can formulate an excluding criterion, i.e. we can specify what kind of spectral 
pattern can never be the result of spurious disturbance from one or more additional grains nearby. 
We also present, as an example, experimental data of this type.

\section{ENERGY DECREMENTS ON TUNNELING}
In our analysis we follow the solution of an orthodox model for a single-grain SET, 
as given in~\cite{Ortho}, with a proper generalisation for a double-grain system.
Whenever possible, we also use the notations introduced in~\cite{Ortho}.

The electrostatic charging energy decrement $\Delta K_{\alpha,\beta}$ 
during the tunneling jump of an electron between the grain $\alpha$ and the
lead electrode $\beta$ can be found directly and equals
(see also Fig.~\ref{ddot} for notations):

\begin{eqnarray}
\Delta \overrightarrow{K}_{\alpha\beta}=\frac{1}{C_{\Delta}^{2}}
\bigg[ &{}&e^2\left(C_{\Sigma \bar{\alpha}}\left((-1)^{\beta} n_\alpha-\frac{1}{2}\right)+C(-1)^{\beta}n_{\bar{\alpha}}\right)+
\nonumber\\
&+&eV(C_{\Sigma \bar{\alpha}}C_{\alpha\bar{\beta}}+CC_{\bar{\alpha}\bar{\beta}})+e\left((-1)^{\beta}V_g+\frac{V}{2}\right)(C_{\Sigma \bar{\alpha}}C_{g\alpha}+CC_{g\bar{\alpha}}) \bigg],
\nonumber\\
\Delta \overleftarrow{K}_{\alpha\beta}=\frac{1}{C_{\Delta}^{2}}
\bigg[ &-&e^2\left(C_{\Sigma \bar{\alpha}}\left((-1)^{\beta}n_\alpha+\frac{1}{2}\right)+C(-1)^{\beta}n_{\bar{\alpha}}\right)-
\nonumber\\
&-&eV(C_{\Sigma \bar{\alpha}}C_{\alpha\bar{\beta}}+CC_{\bar{\alpha}\bar{\beta}})-e\left((-1)^{\beta}V_g+\frac{V}{2}\right)(C_{\Sigma \bar{\alpha}}C_{g\alpha}+CC_{g\bar{\alpha}}) \bigg].
\nonumber\\
\label{DoubleDotK}
\end{eqnarray}
For the tunneling between grains we have

\begin{eqnarray}
\Delta K_{\alpha\bar{\alpha}}=\frac{1}{C_{\Delta}^{2}}
\bigg[-e^2
\left( -C_{\Sigma \bar{\alpha}}\left(n_\alpha-\frac{1}{2}\right)+
C_{\Sigma \alpha}\left(n_{\bar{\alpha}}+\frac{1}{2}\right)+C(n_\alpha-n_{\bar{\alpha}}-1) \right)&+&
\nonumber\\
-e(-1)^{\alpha}V \left( C_{\alpha\alpha}C_{\bar{\alpha}\bar{\alpha}}-C_{\alpha\bar{\alpha}}C_{\bar{\alpha}\alpha}+C_{g\alpha}\frac{C_{\bar{\alpha}\bar{\alpha}}-C_{\bar{\alpha}\alpha}}{2}+C_{g\bar{\alpha}}\frac{C_{\alpha\alpha}-C_{\alpha\bar{\alpha}}}{2} \right)&+&
\nonumber\\
+eV_g\bigg( C_{g\alpha}(C_{\bar{\alpha}\bar{\alpha}}+C_{\bar{\alpha}\alpha})-C_{g\bar{\alpha}}(C_{\alpha\alpha}+C_{\alpha\bar{\alpha}}) \bigg) \bigg].
\nonumber\\
\label{DoubleDotK2}
\end{eqnarray}

Here we defined $C_{\Sigma 1}=C_{11}+C_{12}+C_{g1}+C$, $C_{\Sigma 2}=C_{21}+C_{22}+C_{g2}+C$,
and $C_{\Delta}^{2}=C_{\Sigma 1}C_{\Sigma 2}-C^2$.
We also defined $\bar{\alpha} = 2$ if $\alpha = 1$ and $\bar{\alpha} = 1$ if $\alpha = 2$; 
the same definition for $\bar{\beta}$. 

Equations~(\ref{DoubleDotK}) and~(\ref{DoubleDotK2}) are highly symmetric - see Appendix A.

\section{QUASIPERIODICITY} 
The position of Coulomb staircases in a $\frac{\mathrm dI}{\mathrm dV}(V,V_{g})$ plot 
is determined by the condition that the electrostatic energy decrement on tunneling through some
barrier crosses zero:

$\Delta \overrightarrow{K}_{\alpha \beta}(n_1,n_2,V,V_{g})=0.$ 

From~(\ref{DoubleDotK}) and~(\ref{DoubleDotK2}) we have:
\begin{eqnarray}
V\left( C_{12}+\frac{C}{C_{\Sigma2}}C_{22} \right)-
\left( V_{g}-\frac{V}{2} \right)\left( C_{g1}+\frac{C}{C_{\Sigma2}}C_{g2} \right)
&=&e\left( n_1+\frac{C}{C_{\Sigma2}}n_2 \pm \frac{1}{2} \right),
\nonumber\\
-V\left( C_{11}+\frac{C}{C_{\Sigma2}}C_{21} \right)-
\left( V_{g}+\frac{V}{2} \right)\left( C_{g1}+\frac{C}{C_{\Sigma2}}C_{g2} \right)
&=&e\left( n_1+\frac{C}{C_{\Sigma2}}n_2 \pm \frac{1}{2} \right),
\nonumber\\
V\left( \frac{C}{C_{\Sigma1}}C_{12}+C_{22} \right)-
\left( V_{g}-\frac{V}{2} \right)\left( \frac{C}{C_{\Sigma1}}C_{g1}+C_{g2} \right)
&=&e\left( \frac{C}{C_{\Sigma1}}n_1+n_2 \pm \frac{1}{2} \right),
\nonumber\\
-V\left( \frac{C}{C_{\Sigma1}}C_{11}+C_{21} \right)-
\left( V_{g}+\frac{V}{2} \right)\left( \frac{C}{C_{\Sigma1}}C_{g1}+C_{g2} \right)
&=&e\left( \frac{C}{C_{\Sigma1}}n_1+n_2 \pm \frac{1}{2} \right),
\nonumber\\
V\bigg( C_{11}C_{22}-C_{12}C_{21}+C_{g1}\frac{C_{22}-C_{21}}{2} &+&C_{g2}\frac{C_{11}-C_{12}}{2}\bigg)+
\nonumber\\
+ V_g \bigg ( C_{g1}(C_{22}+C_{21})&-&C_{g2}(C_{11}+C_{12}) \bigg) =
\nonumber\\
=e\bigg( -(C_{\Sigma 2}-C)n_1 +(C_{\Sigma 1}-C) n_2 \pm (\frac{1}{2}C_{\Sigma 1} + \frac{1}{2}C_{\Sigma 2} &-&C) \bigg).
\label{Slops}
\end{eqnarray} 

Because of time-reversal symmetry between tunneling jumps $\alpha \to \beta$ and $\beta \to \alpha$,
the conditions $\Delta \overrightarrow{K}_{\alpha \beta}(n_{\alpha})=0$ and
$\Delta \overleftarrow{K}_{\alpha \beta}(n_{\alpha}+1)=0$
impose the same relationship on $V, V_g$, and only 5 distinct sets of pecularities are defined by~(\ref{Slops}).

We can draw some important conclusions from~(\ref{Slops}):

{\bf a.} All Coulomb pecularities form straight lines in a 
$\frac{\mathrm dI}{\mathrm dV}(V,V_{g})$ plot.
Indeed, any equation in~(\ref{Slops}) has a linear form
\begin{equation}
A_{\alpha\beta}V+B_{\alpha \beta}V_{g}=C_{\alpha \beta}n_1+D_{\alpha \beta}n_2+const.
\label{Linearity}
\end{equation} 

{\bf b.} Each tunneling barrier defines a set of pecularities which are parallel
to each other, i. e. a double dot system has maximum five distinct slopes in the $\frac {\mathrm dI}{\mathrm dV}$ pattern.   

{\bf c.} There is no more periodicity with respect to a gate
voltage, like in a single-dot system.
Nevertheless, the whole $\frac {\mathrm dI}{\mathrm dV}$ pattern is quasiperiodic, because
all staircases coming from the first grain have offsets $n_1+(C/C_{\Sigma2})n_2+\cdots$,
from the second one -  $(C/C_{\Sigma1})n_1+n_2+\dots$, and from intergrain tunneling -
$(C_{\Sigma 2}-C)n_1 -(C_{\Sigma 1}-C)n_2+\cdots$.

\section{MASTER EQUATION FOR EVOLUTION}
If both grains have a continuos spectrum, then for
tunneling rates $\Gamma_{\alpha,\beta}(n_1,n_2)$ through the lead-grain gaps we have:
\begin{eqnarray}
\overrightarrow{\Gamma}_{\alpha\beta}&=&\frac{\frac{1}{e}I_{\alpha\beta}\left(\frac{\Delta \overrightarrow{K}_{\alpha\beta}}{e}\right)}
{1-\exp\left(-\frac{\Delta \overrightarrow{K}_{\alpha\beta}}{kT}\right)},
\nonumber\\
\overleftarrow{\Gamma}_{\alpha\beta}&=&\frac{\frac{1}{e}I_{\alpha\beta}\left(\frac{\Delta \overleftarrow{K}_{\alpha\beta}}{e}\right)}
{1-\exp\left(-\frac{\Delta \overleftarrow{K}_{\alpha\beta}}{kT}\right)},
{\rm \; \alpha, \beta = 1,2;}
\label{Gammas2}
\end{eqnarray}
and for intergrain tunneling rates:
\begin{equation}
\Gamma_{\alpha\bar{\alpha}}=\frac{\frac{1}{e}I_{\alpha\bar{\alpha}}\Big(\frac{\Delta K_{\alpha\bar{\alpha}}}{e}\Big)}
{1-\exp\Big(-\frac{\Delta K_{\alpha\bar{\alpha}}}{kT}\Big)}, {\rm \; \alpha = 1,2,}
\end{equation}
where $I_{\alpha\beta}(V)$ is the current-voltage characteristics of the corresponding tunneling barrier.
Assuming Ohms law, we put $I_{\alpha\beta}(V) = G_{\alpha\beta}V$, but all formulas can be easily 
generalized for the case of a discrete energy spectrum in the grains.

The master equation for the double dot system is
\begin{eqnarray}
\dot{p}_{n1,n2}= -\Big(\overrightarrow{\Gamma}_{11}(n_1,n_2)&+&\overleftarrow{\Gamma}_{11}(n_1,n_2)+
               \overrightarrow{\Gamma}_{12}(n_1,n_2)+\overleftarrow{\Gamma}_{12}(n_1,n_2)+
               \nonumber\\
               \overrightarrow{\Gamma}_{21}(n_1,n_2)&+&\overleftarrow{\Gamma}_{21}(n_1,n_2)+
               \overrightarrow{\Gamma}_{22}(n_1,n_2)+\overleftarrow{\Gamma}_{22}(n_1,n_2)+
               \nonumber\\
               +\Gamma_{12}(n_1,n_2)&+&\Gamma_{21}(n_1,n_2)\Big)p_{n_1,n_2}+
               \nonumber\\
               +\Big( \overrightarrow{\Gamma}_{11}(n_1-1,n_2)&+&\overleftarrow{\Gamma}_{12}(n_1-1,n_2)\Big)p_{n_1-1,n_2}+
               \nonumber\\
               +\Big( \overrightarrow{\Gamma}_{21}(n_1,n_2-1)&+&\overleftarrow{\Gamma}_{22}(n_1,n_2-1)\Big)p_{n_1,n_2-1}+
               \nonumber\\
                +\Big( \overleftarrow{\Gamma}_{11}(n_1+1,n_2)&+&\overrightarrow{\Gamma}_{12}(n_1+1,n_2)\Big)p_{n_1+1,n_2}+
               \nonumber\\
                +\Big( \overleftarrow{\Gamma}_{21}(n_1,n_2+1)&+&\overrightarrow{\Gamma}_{22}(n_1,n_2+1)\Big)p_{n_1,n_2+1}+
               \nonumber\\
               +\Gamma_{21}(n_1-1,n_2+1)p_{n_1-1,n_2+1}&+&\Gamma_{12}(n_1+1,n_2-1)p_{n_1+1,n_2-1}.
\label{Master2}
\end{eqnarray}

And the current is given by 
\begin{eqnarray}
I=e\sum\limits_{n1,n2=-\infty}^{+\infty} \Big(
 \overrightarrow{\Gamma}_{11}(n_{1},n_{2})-\overleftarrow{\Gamma}_{11}(n_{1},n_{2})+ 
 \overrightarrow{\Gamma}_{21}(n_{1},n_{2})-\overleftarrow{\Gamma}_{21}(n_{1},n_{2})
\Big)p_{n_1,n_2}.
\nonumber\\
\label{DoubleDotCurrent}
\end{eqnarray}

\section{NUMERIC SOLUTION OF A DOUBLE DOT SET}
Master equation (\ref{Master2}) can be solved numerically in a straightforward manner:
for any given $(V, V_g)$ one should trace the evolution of $p_{n,m}$ 
until an equilibrium distribution is reached 
(see Appendix B for the details of the numeric procedure), 
the current is then given by (\ref{DoubleDotCurrent}).
This way the current can be computed for an {\it arbitrary} double grain system.

To test the program, we solved a general double dot system and compared the program output
with the result of a Monte-Carlo simulation obtained with a commercial single electron circuit simulation
software - SIMON 2.0~\cite{Simon}.
As one can see, all Coulomb peculiarities  have identical positions in both
Fig.~\ref{SimulT}a and Fig.~\ref{SimulT}b.

\section{DOUBLE DOT SYSTEM WITH STRONGLY ASYMMETRIC TUNNELING BARRIERS}

Fig.~\ref{SimulT} illustrates the common rule: in a multigrain system any tunneling gap 
gives a set of peculiarities on $\frac{\mathrm dI}{\mathrm dV}(V,V_{g})$ plot 
with one and just one unique slope.
Equation set (\ref{Slops}), however, only specifies possible staircase positions, not amplitudes. 
The amplitudes depend on probability distribution $p_{n_1,n_2}$ as defined by (\ref{Master2}).

For example, lets consider the case of negative bias voltage $V$ with electrons flowing in Source-to-Drain direction. 
Whenever the small variation of $V$ and/or $V_g$ makes, say, 
$\Delta \overrightarrow{K}_{11}(n,m)$ positive, electron from Fermi level in a Source lead can tunnel 
into a Fermi level in the first grain.
One additional intermediate charge state $\{n+1,m\}$ becomes accessible, resulting in one extra term in current
sum (\ref{DoubleDotCurrent}) due to to $\{n,m\} \rightarrow \{n+1,m\}$ transitions.
The resulting current increment, however, is proportional to the probability of a system to be found in a $\{n,m\}$ 
state multiplied by $\{n,m\} \rightarrow \{n+1,m\}$ tunneling rate:
$\overrightarrow{\Gamma}_{11}(n,m) p_{n,m}$.
If $p_{n,m}$ is small, i.e. the $\{n,m\}$ state is statistically unfeasible, 
then a newly opened channel does not contribute much the total current, 
and the corresponding staircase is suppressed.

For clarification, we can refer to a single grain SET.
In a symmetric case $(G_1=G_2)$ all charge states, accessible from the energy considerations, 
have comparable feasibility: $p_{n_{max}} \sim p_{n_{max}-1} \sim \dots  \sim p_{-n_{max}}$.
On the contrary, in the asymmetric case $G_1 \gg G_2$ the grain is almost in equilibrium with the 
low resistance electrode.
In a rare event of tunneling through the low-conductive gap $G_2$ equilibrium is violated, 
but almost immediately restored by tunneling through a high-conductive barrier $G_1$. 
Two successive tunneling events through $G_2$ not interleaved by tunneling through $G_1$ are even less feasible.
As a result, the charge fluctuations are strongly suppressed and in the $G_2 \rightarrow 0$ limit  
the probability distribution 
reduces to a single charge state $n_{max}$: $p_n= \delta_{n, n_{max}}$.

Whenever a bias or gate increment makes tunneling through the $G_1$ gap into the $n+1$ state possible,
an equilibrium grain charge switches to $e(n+1)$, and the current is incremented stepwise.
On the contrary, when the voltage variation opens a new tunneling channel through the $G_2$ gap,
the equilibrium charge remains the same, and no current step takes place.
While a symmetric SET has two distinct staircase families, the first one defined by
$K_1(n,V,V_g) = 0$ and the second by $K_2(n,V,V_g) = 0$, 
in an asymmetric system one of them is suppressed.

Exactly the same kinetic effect takes place for a double grain (multigrain) SET:
if tunneling gaps have strongly different conductances $G_{\alpha\beta}$, 
then the vast majority of accessible charge states becomes unfeasible, and for any given pair $(V,V_g)$ 
just one charge state $\{n_1(V,V_g),n_2(V,V_g)\}$ dominates.
If only $V$ and/or $V_g$ variation changes the dominant charge state, a prominent staircase appears.
For any grain $\alpha$, the grain charge is defined by the most conducting tunneling gap $\beta_{\mathrm {max}}$, 
and the charge state changes when $K_{\alpha \beta_{\mathrm {max}}}(n_1,n_2,V,V_g)$ crosses zero. 
All other staircases, corresponding $K_{\alpha, \beta \ne \beta{\mathrm {max}}}$ are supressed.

To summarize, in a multigrain system each tunneling gap defines one staircase family with a distinct 
slope in the $(V,V_g)$ plane. But in a strongly asymmetric system only one staircase family per every grain 
survives - the one corresponding to the most conductive link.
Although in the double grain system there may be up to five distinct staircase families, 
in the asymmetric case only two will stay - the first one corresponding to the $n_1$ increments and 
the second one to $n_2$ increments. Three-grain system will have at least three staircase families and so on.
And vice versa - a system with $n$ distinct staircase slopes has no more than $n$ grains.

This qualitative consideration is confirmed by the numeric result presented in Fig.~\ref{SimAntSy}.

\section{A SYSTEM WITH TWO STAIRCASE FAMILIES: POSSIBLE TOPOLOGIES}
As it was discussed above, a system with two staircase families may have maximum two grains. 
Each grain must have one strong tunneling link with conductance much higher than the other 
tunneling gaps adjacent to this grain. This allows us to list all possible topologies directly.
We note first that each grain has three links, and any one could be dominant. 
Therefore, for two grains we have $3 \times 3$ different possibilities, all shown in Fig~\ref{Topolog2}.   
(note that for generality we assume that the grains have different size).
Variants {\bf a}, {\bf c}, {\bf e}, and {\bf g} are obviously symmetric to {\bf b}, {\bf d}, {\bf f}, and {\bf h} 
respectively. Variant {\bf i} has no more than two staircase families if only one 
of the grain-electrode gaps has much higher conductance than the others. In this case 
the system reduces either to {\bf d} or {\bf f}. To summarize, we have four distinct cases: 
{\bf b}, {\bf d}, {\bf f}, and {\bf h}; we shall consider each one in more detail below\footnote{
Note that it is natural to assume in {\bf b}, {\bf d}, and {\bf f} cases that a smaller 
grain is closer to the left electrode, and the bigger grain - to the right one. 
That's why all numeric results presented below were computed for a system with $C_{11} > C_{12}$ 
and $C_{21} < C_{22}$. We put this constraint to make the staircase slopes more realistic, ¨
it does not affect the cardinal pattern of dI/dV plot.}.

Grains are linked to different electrodes - the case in Fig.~\ref{Topolog2}b. 
Two visible staircase families are those coming from the gaps $G_{11}$ and $G_{22}$.
The ${\mathrm d}I/{\mathrm d}V(V,V_g)$ plot looks slightly different depending on which 
of the remaining tunneling gaps - $G_{12}$, $G_{21}$ or $G$ has the highest conductivity.
The current either flows mainly through the big grain ($G_{21} > \mathrm{max}(G, G_{12})$ - 
the numeric result is shown in Fig.~\ref{ddotall1}b1), 
through the small one ($G_{12} > \mathrm{max}(G, G_{21})$ - see Fig.~\ref{ddotall1}b2), 
or two grains connected 
in series ($G > \mathrm{max}(G_{21}, G_{12})$ - presented in Fig.~\ref{SimAntSy}b). 
Note that in all cases the 
Coulomb staircases have exactly the same position in the $(V,V_g)$ plane, but the 
staircase amplitudes (and even the signs) are different.
The regions of the Coulomb blockade are also different. In the general case of $G_{12} 
\sim G_{21} \sim G$ the Coulomb blockade is lifted as soon as at least one of these 
current channels opens for transport.

A big grain is linked to a small one, which is linked to an electrode - the case 
in Fig.~\ref{Topolog2}d.
The current either goes through the big grain ($G_{22}>G_{12}$) or 
bypasses it ($G_{22}<G_{12}$ ). The corresponding ${\mathrm d}I/{\mathrm d}V$ 
plots are shown in Fig.~\ref{ddotall1}d1 and Fig.~\ref{ddotall1}d1 respectively.

A small grain is linked to a big one, which is linked to an electrode - the case in Fig.~\ref{Topolog2}f. 
Two possible topologies complementary to d1 and d2 are presented in 
Fig.~\ref{ddotall1}f1 and Fig.~\ref{ddotall1}f2 together with ${\mathrm d}I/{\mathrm d}V$ plots.

The last possibility - both grains are linked to the same electrode - is shown in Fig.~\ref{Topolog2}h.
If the current goes mainly through the small grain, we have the ${\mathrm d}I/{\mathrm d}V$ 
pattern presented in Fig.~\ref{ddotall1}h1, otherwise - the one in Fig~\ref{ddotall1}h2.

In any case ${\mathrm d}I/{\mathrm d}V$ pattern has a simple structure: there are big areas 
shaded by parallel and equdistant spectral lines 
(one is shown by dotted line in Fig.~\ref{ddotall1}h1).
The reason for this highly regular structure is quite obvious for topologies in b1, b2, h1, h2 - 
a single grain SET coupled to a Single Electron Box (SEB).
Indeed, until SEB charge $Q_{\mathrm{SEB}}$ remains constant, the SEB affects SET 
in essentially the same way as a constant background charge (of course, there is also 
some capacitance renormalisation). 
So all SET staircases are periodic with $(V,V_g)$. 
Whenever $Q_{\mathrm{SEB}}$ changes, the SET feels a different "background charge" and the phase of its 
Coulomb blockade oscillations is shifted.
Remarkably, this high regularity survives even if two grains are connected in series, as in d1 and f1. 

\section{COMPARISON WITH EXPERIMENT}
The analysis presented in this paper was to a large extent motivated by our recent experiments 
with nanoconstrictions in Quench-Condenced (QC) films~\cite{OurAPL,OurJLTP}.
We have developed a method to define constrictions with width and length of about 5~nm, bridged 
to two macroscopic electrodes.
At small thiknesses QC films are granular~\cite{Valles}, so there are typically a few metallic 
grains separated by tunneling gaps within such a constriction - see Fig.~\ref{Sample} for the sample geometry.
By additional {\it in situ} film depositions we merged some clusters together, finally reducing the 
sample topology to a single grain SET.
The only opportunity to follow the sample topology was the analysis of the sample's 
$\mathrm{d}I/\mathrm{d}V(V,V_g)$ pattern.
We used the technique presented here 
to determine the number of grains in the constriction, to reconstruct 
the sample topology, and to extract all grain capacitances.
The fast computation time (compared to the Monte-Carlo method) makes this procedure practical. 

The strong exponential dependence of the tunneling conductance on the gap width greatly 
simplifies the spectral pattern. 
Indeed, we can assume that the tunneling 
gaps between the grains are randomly and uniformly 
distributed from 1~\AA~to 10~\AA. 
Then the tunneling resistances are randomly and uniformly distributed in 
{\it logarithmic} scale from 10~k${\Omega}$ and up to 1~G${\Omega}$. 
This virtually eliminates the kinetic charge fluctuations, and washes out all 
but one staircase family per grain, i. e. every staircase family identifies the 
unique grain.

A typical experimental data are shown in Fig.~\ref{FitDemo} together with a numeric fit.
The reconstructed sample topology is presented in Fig.~\ref{FitDemo}c: two grains, with charging energies 
$e^2/2C_\Sigma$ about 9 and 34~meV, are connected in series; and there is a small 
single electron box with charging energy $\sim 52$~meV nearby.

Note that the inter-grain capacitance $C$ is always small compared to coupling to the leads or 
the gate electrode (see caption to Fig.~\ref{FitDemo}).
This is because an inter-grain interaction is effectively screened by the common gate electrode.
For the sample in Fig.~\ref{FitDemo} the gate oxide layer was about 3~nm thick, so for two 
grains bigger than $\sim 3$~nm we always have $C < C_g$. Moreover, any two grains 
placed more than a few nanometers away virtually do not interfere.
This allows to apply our 
analysis of a {\it double-grain} system recursively, and to identify all grains 
in a {\it multi-grain} system one by one.

Whenever an additional deposition merges two grains, one staircase family disappears.
This evolution eventually leads to a single grain SET, and after a few additional depositions, 
to a single tunneling gap separating two macroscopic electrodes~\cite{OurJLTP}.

Intriguingly, for samples with very high charging energies ($\gtrsim 100$~meV), we have always 
observed some
non-orthodox spectral lines like the ones marked with $\gamma$ in Fig.~\ref{Exper}.
All lines $\gamma$ have the same slope, different from the slope $\beta$ of the main Coulomb staircases.
As it was shown previously, a system with two distinct slopes on the ${\mathrm d}I/{\mathrm d}V$ pattern 
has no more than two grains involved in current trunsport. Would the lines $\gamma$ be due to recharging 
of some other grains, they all must come from recharging of one and the same grain. 
In this case they must be equidistant, which obviously does not take place.

\section{CONCLUSION}
We proved the following statements for a multigrain system in the framework of the ortodox model: 

\begin{enumerate}
\item
All Coulomb peculiarities in differential conductance
form straight lines on the gate-bias plane.

\item
Each tunneling gap defines a set of staircases, which are parallel to each other. 
In an asymmetric system with strongly different tunneling resistances only one distinct staircase 
family per every grain survives - the one corresponding to the most conductive link; 
all other staircases are suppressed.

\item
There may be a maximum of $n$ grains in a system with $n$ staircase slopes.
\end{enumerate}

We presented the exact solution for an arbitrarily connected double-grain system.
We demonstrated that although the zero-bias conductance is stochastic~\cite{Ruzin2}, 
the large-scale spectroscopy pattern ${\mathrm d}I/{\mathrm d}V(V, V_g)$ 
(differential conductance as a function of bias and gate voltages) is regular - 
all Coulomb staircases are quasiperiodic.

We listed all possible topologies for a system with two different staircase slopes and presented the 
corresponding spectral patterns.

Finally, we proved the following criterion:

If there are only two slopes $\alpha$ and $\beta$ for all spectral lines on spectroscopy 
pattern, and the lines with the slope $\beta$ are {\it not equidistant} within the regions between two adjacent lines $alpha$, 
then they can not be attributed 
to the presence of one or more additional grains in the 
sample\footnote{This criterion can be generalized in a straightforward manner to the case of $n$ slopes on
 ${\mathrm d}I/{\mathrm d}V$ pattern, but this situation seems to be of low practical importance.
}.

To conclude, we have analysed a multigrain system with a shared gate and found that the 
sample topology can be unambiguously deduced from ${\mathrm d}I/{\mathrm d}V(V, V_g)$ pattern.
We demonstrated how to distinguish peculiarities in tunneling spectra of single-grain 
samples from possible disturbance caused by some extra grains.
Presented analysis can be applied for numerous systems, where the experimentalist is 
not able to image the sample directly. 

\section{ACKNOWLEDGMENTS}
We are grateful to T. Claeson and A.Ya. Tzalenchuk 
for illuminating discussions and critical readings of the manuscript.
The work was supported by 
the Royal Swedish Academy of Sciences, through its Nobel Institute,
by Swedish Foundation of Strategic Research,
and by European Union under IST-FET NANOMOL program. 

\section{APPENDIX A: SYMMETRY CONSIDERATIONS}
Equations~(\ref{DoubleDotK}) and~(\ref{DoubleDotK2}) are invariant with respect to the following symmetry
transformations.

{\bf a.} We can swap indices marking the grain numbers, i. e. we can redeclare grain no 1 as 2 and
vice versa. So as a result of simultaneous replacement 

$n_{1} \leftrightarrows  n_{2}, \, C_{11} \leftrightarrows C_{21}, \,
C_{12} \leftrightarrows C_{22}, \, C_{g1} \leftrightarrows C_{g2}, \,
\rm and \; C_{\Sigma 1} \leftrightarrows C_{\Sigma 2}$ \\
we will have the following transformations:

$\Delta \overrightarrow{K}_{1 \beta}(n_{1},n_{2}) \leftrightarrows \Delta \overrightarrow{K}_{2 \beta}(n_{2},n_{1}), \,
\Delta \overleftarrow{K}_{1 \beta}(n_{1},n_{2}) \leftrightarrows \Delta \overleftarrow{K}_{2 \beta}(n_{2},n_{1}), \, \rm and \, \\
\Delta K_{12}(n_{1},n_{2}) \leftrightarrows \Delta K_{21}(n_{2},n_{1})$.

{\bf b.} Left/right symmetry: we can flip the whole network horizontally with simultaneous
inversion of the bias voltage, which is equivalent to the following substitution:

$C_{11} \leftrightarrows C_{12}, \, C_{21} \leftrightarrows C_{22}, \, V \leftrightarrows -V.$\\
As a result, the tunneling jump from, say, the left tunneling barrier will be transformed
into a jump through the right barrier in the {\it opposite} direction, so as

$\Delta \overrightarrow{K}_{\alpha 1}(V,V_{g}) \leftrightarrows \Delta \overleftarrow{K}_{\alpha 2}(-V,V_{g}), \,
\Delta \overleftarrow{K}_{\alpha 1}(V,V_{g}) \leftrightarrows \Delta \overrightarrow{K}_{\alpha 2}(-V,V_{g}). \, \\
$

Energy decrements for intergrain tunneling $\Delta K_{12}$ and $\Delta K_{21}$ are not affected by this
transformation and they will convert to themselves.

{\bf c.} Time inversion symmetry: if for any tunneling event from state $n_{1}, n_{2}$ to state $n_{1}', n_{2}'$
the energy decrement is $\Delta K$, then the inverse tunneling process from $n_{1}', n_{2}'$ to $n_{1}, n_{2}$ must have
the opposite energy decrement $-\Delta K$, and so\\
$\Delta \overrightarrow{K}_{11}(n_{1},n_{2}) = -\Delta \overleftarrow{K}_{11}(n_{1}+1,n_{2}), \,
\Delta \overrightarrow{K}_{12}(n_{1},n_{2}) = -\Delta \overleftarrow{K}_{12}(n_{1}-1,n_{2}), \, \\
\Delta \overrightarrow{K}_{21}(n_{1},n_{2}) = -\Delta \overleftarrow{K}_{21}(n_{1},n_{2}+1), \, 
\Delta \overrightarrow{K}_{22}(n_{1},n_{2}) = -\Delta \overleftarrow{K}_{22}(n_{1},n_{2}-1), \, \\
{\mathrm{and}} \; \Delta K_{12}(n_{1},n_{2}) = -\Delta K_{21}(n_{1}-1,n_{2}+1).$

Any of {\bf a}-{\bf c} symmetries reduces the number of independent formulas
in~(\ref{DoubleDotK}) by a factor of 2, so there is really just one independent equation
in~(\ref{DoubleDotK}). 

There is yet another symmetry in~(\ref{DoubleDotK})
and~(\ref{DoubleDotK2}):

{\bf d.} All energy decrements change the sign and the tunneling direction on simultaneous
inversion of all voltages and charges, i. e. 

$\Delta \overrightarrow{K}_{\alpha \beta}(n_{1}, n_{2}, V, V_{g})= -\Delta \overleftarrow{K}_{\alpha \beta}(-n_{1}, -n_{2}, -V, -V_{g}), \, \rm and 
$

$   
\Delta \overleftarrow{K}_{\alpha \beta}(n_{1}, n_{2}, V, V_{g})= -\Delta \overrightarrow{K}_{\alpha \beta}(-n_{1}, -n_{2}, -V, -V_{g}),
$

If we apply this transformation to the expression for current~(\ref{DoubleDotCurrent}) then the current will change
sign:
\begin{equation}
I(V,V_{g})=-I(-V,-V_{g}), \, {\mathrm {and}} \, \, \frac{\mathrm dI}{\mathrm dV}(V,V_{g})=\frac{\mathrm dI}{\mathrm dV}I(-V,-V_{g}).
\end{equation}

So the $I(V,V_{g})$ plot is antisymmetric, and the $\frac{\mathrm dI}{\mathrm dV}(V,V_{g})$ plot is symmetric.
In the presence of a background charge, the relative phase of Coulomb oscillations on the first and the second grain will be shifted and this symmetry will be violated.
 
If we complete transformation {\bf d} with the inversion of the electron charge $e \to -e$ then 
the current will not change. This could be reformulated as follows: the current through the system
does not depend on the sign of charge carriers, i. e. whether the current transport is carried by
electrons or holes.

\section{APPENDIX B: NUMERICAL RECIPIES}
The master equation~(\ref{Master2}) is a system of linear differential equations. If we present
the probability distribution $p_{n_{1},n_{2}}$ as a vector $\vec p = \vert p_{j} \rangle$ where $j$ is
an index running over all $(n_{1},n_{2})$ pairs, then we can rewrite~(\ref{Master2}) as 
\begin{equation}
\dot {\vec p} = -\widehat{\Gamma} \vec p.
\label{MasterVec}
\end{equation}

The formal solution of~(\ref{MasterVec}) is
\begin{eqnarray}
\vec p(t) &=& e^{-t\widehat{\Gamma}} \vec p(0) = (
\mathrm {\widehat E} - t \widehat \Gamma + \frac{t^2}{2} \widehat \Gamma^2
-\frac{t^3}{6} \widehat \Gamma^3+\frac{t^4}{24} \widehat \Gamma^4+\cdots)
\nonumber\\
&=& (\mathrm {\widehat E} - t \widehat \Gamma)(\mathrm {\widehat E} - \frac{t}{2} \widehat \Gamma)
(\mathrm {\widehat E} - \frac{t}{3} \widehat \Gamma)(\mathrm {\widehat E} - \frac{t}{4} \widehat \Gamma)\vec p(0)
+o(t^4),
\label{Expansion}
\end{eqnarray}
where $\mathrm {\widehat E}$ is the unity matrix.

As one can see from~(\ref{Expansion}), the standard fourth-order Runge-Kutta method reduces
in our case to four recursive calls to a first-order Eiler's extrapolation formula with steps $\frac{t}{4}$,
$\frac{t}{3}$,  $\frac{t}{2}$, $t$.
 
To minimise execution time, it is better to use an adaptive step $t$. Ideally, it should be
inversely proportional to the maximum eigenvalue of the evolution matrix $\widehat \Gamma$ $\lambda_{max}$.
Practically, it is good enough to replace an unknown $\lambda_{max}$ with the sum of all eigenvalues
$\sum
 \lambda_{j} = tr(\widehat \Gamma)$, where the trace of the evolution matrix can be found
from~(\ref{Master2}) and equals
\begin{equation}
tr(\widehat \Gamma)= \sum \limits_{n1,n2 = -\infty}^{\infty}
\Big(
\sum \limits_{\alpha, \beta = 1,2}
(\overrightarrow{\Gamma}_{\alpha \beta}(n_{1},n_{2})+\overleftarrow{\Gamma}_{\alpha \beta}(n_{1},n_{2}))+
\sum \limits_{\alpha = 1,2}
\Gamma_{\alpha \bar{\alpha}}(n_{1},n_{2})
\Big).
\end{equation}

}


\begin{figure}[!t]
\begin{center}
\includegraphics[width=3in]{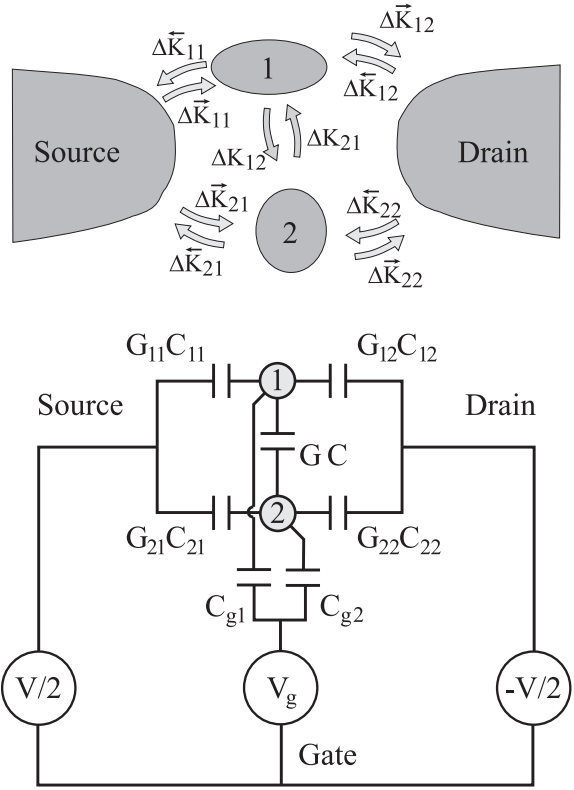}
\end{center}
\caption{Top: system topology and energy decrements on tunneling jumps; 
Bottom: equivalent schematics.}
\label{ddot}
\end{figure}

\begin{figure}[!hbt]
\begin{center}\leavevmode
\includegraphics[width=3in]{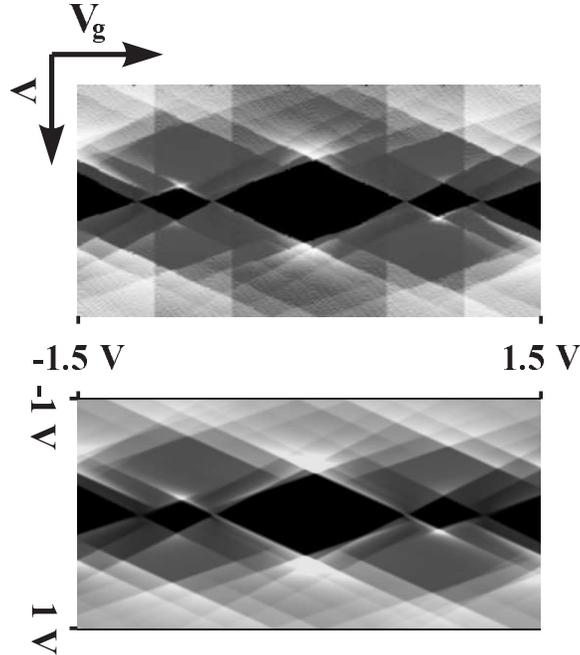}
\caption{General case of a double dot SET: 
differential conductance plot as a function of bias and gate voltages. 
{\bf a.}- The result of Monte-Carlo simulation.
{\bf b.}- Direct numeric integration of master equation.
System parameters:
$C_{11}=16 \times 10^{-20}$~F, $C_{12}=8 \times 10^{-20}$~F,
$C_{21}=9 \times 10^{-20}$~F, $C_{22}=13 \times 10^{-20}$~F,
$C_{g1}=8 \times 10^{-20}$~F, $C_{g2}=11 \times 10^{-20}$~F,
$C=10 \times 10^{-20}$~F,
$R_{11}=3 \times 10^{9}$~$\Omega$, $R_{12}=2 \times 10^{9}$~$\Omega$,
$R_{21}=1 \times 10^{9}$~$\Omega$, $R_{22}=2 \times 10^{9}$~$\Omega$,
$R=20 \times 10^{9}$~$\Omega$.}
\label{SimulT}
\end{center} 
\end{figure}

\begin{figure}[!hbt]
\begin{center}\leavevmode
\includegraphics[width=3in]{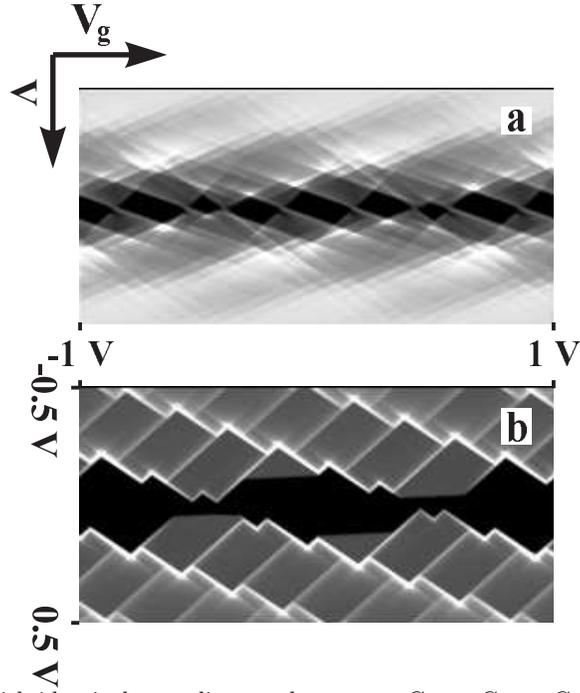}
\caption{{\bf a.}- Double dot system with identical tunneling conductances: 
$G_{11}=G_{12}=G_{21}=G_{22}=G$.
There are five distinct staircase slopes.
{\bf b.}- Every grain has one well-conducting tunneling gap which governs the grain charge: 
$G_{11}=G_{22}=30G$; $G_{12}=G_{21}=0$.
All but two staircase families are suppressed.
Same capacitors for symmetric and asymmetric systems:
$C_{11}=45 \times 10^{-20}$~F, $C_{12}=6 \times 10^{-20}$~F,
$C_{21}=18 \times 10^{-20}$~F, $C_{22}=126 \times 10^{-20}$~F,
$C_{g1}=19 \times 10^{-20}$~F, $C_{g2}=47 \times 10^{-20}$~F,
$C=37 \times 10^{-20}$~F.
}
\label{SimAntSy}
\end{center} 
\end{figure}

\begin{figure}[hbt]
\begin{center}\leavevmode
\includegraphics[width=2in]{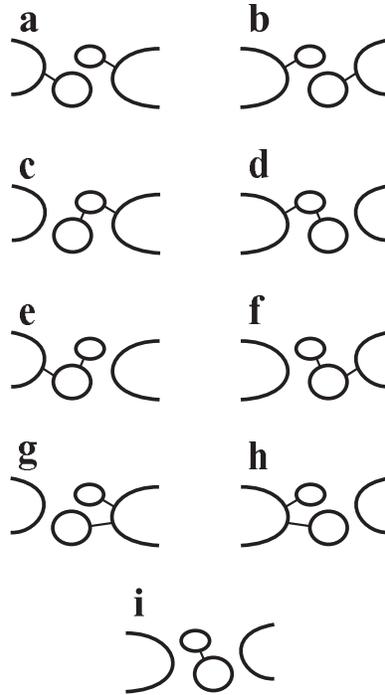}
\caption{All possible topologies for a double grain system with no more than two staircase families.
The case {\bf a} produces two families, associated with tunneling gaps $G_{21}$ and $G_{12}$, 
the case {\bf b} - families associated with $G_{11}$ and $G_{22}$, and so on.
}
\label{Topolog2}
\end{center} 
\end{figure}

\begin{figure}[!hbt]
\begin{center}\leavevmode
\includegraphics[width=6in]{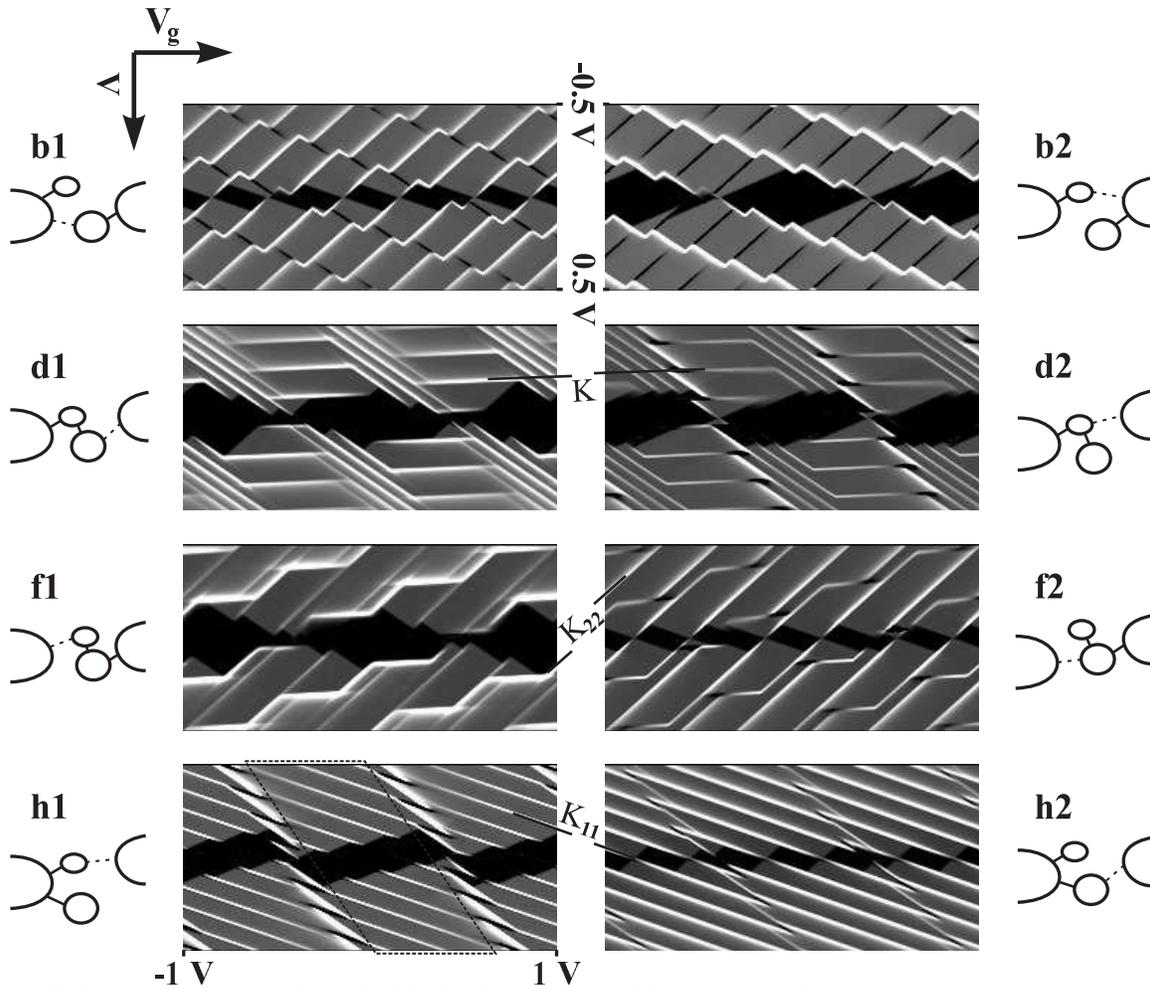}
\caption{
$\mathrm{d}I/\mathrm{d}V(V,V_g)$ patterns for a double-dot SET: all possible cases (together with Fig.~\ref{SimAntSy}b) 
when just two staircase families are present. 
For any topology sketch one solid line indicates the most conducting link for the bigger grain and another 
solid line - the most conducting link for the smaller grain.
The dotted line shows the main current channel.
Note that in cases {\bf b1} and {\bf b2} the second grain gives the staircases with a negative 
differential conductance. There are also small areas of negative 
${\mathrm d}I/{\mathrm d}V$ in {\bf d2}, {\bf f2}, {\bf h1}, and {\bf h2}.
}
\label{ddotall1}
\end{center} 
\end{figure}

\begin{figure}[!b]
\begin{center}\leavevmode
\includegraphics[width=2in]{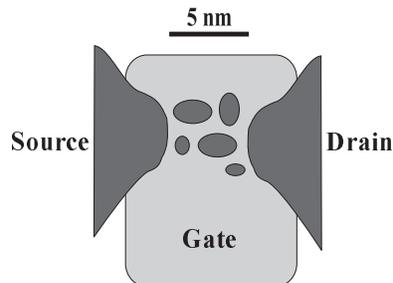}
\caption{ 
Sample geometry: Thick low resistive electrodes are bridged by $5 \times 5$~$nm^2$ network of 
metallic clusters. Beneath is an electrostatic gate electrode made of the oxidized Aluminium.
}
\label{Sample}
\end{center} 
\end{figure}

\begin{figure}[!b]
\begin{center}\leavevmode
\includegraphics[width=3in]{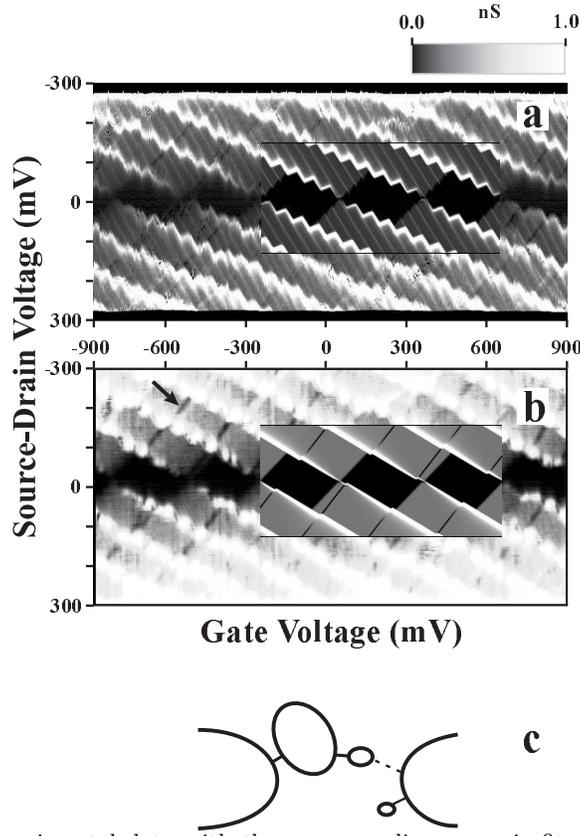}
\caption{A typical example of experimental data with the corresponding numeric fit.
{\bf a.}-
Two grains connected in series. The widest tunneling gap is between 
the small grain and the electrode - the same topology as in Fig.~\ref{ddotall1}f1.
Inset: a numeric fit with parameters
$G_{21}=G=30G_{12}$; $G_{11}=G_{22}=0$;
$C_{11}=8 \times 10^{-20}$~F, $C_{12}=134 \times 10^{-20}$~F,
$C_{21}=380 \times 10^{-20}$~F, $C_{22}=114 \times 10^{-20}$~F,
$C_{g1}=53 \times 10^{-20}$~F, $C_{g2}=352 \times 10^{-20}$~F,
$C=41 \times 10^{-20}$~F.
{\bf b.}-
The same sample as in {\bf a}. Current gradient across the direction indicated
by an arrow was computed to highlight the third staircase family almost unresolvable
in {\bf a}. The third grain is coupled the same way as in Fig.~\ref{ddotall1}b2.  
Inset: a numeric fit with parameters
$G_{11}=G_{22}=30G_{12}$; $G_{21}=G=0$;
$C_{11}=45 \times 10^{-20}$~F, $C_{12}=120 \times 10^{-20}$~F,
$C_{21}=37 \times 10^{-20}$~F, $C_{22}=44 \times 10^{-20}$~F,
$C_{g1}=53 \times 10^{-20}$~F, $C_{g2}=65 \times 10^{-20}$~F,
$C=7 \times 10^{-20}$~F.
}
\label{FitDemo}
\end{center} 
\end{figure}

\begin{figure}[!b]
\begin{center}\leavevmode
\includegraphics[width=3in]{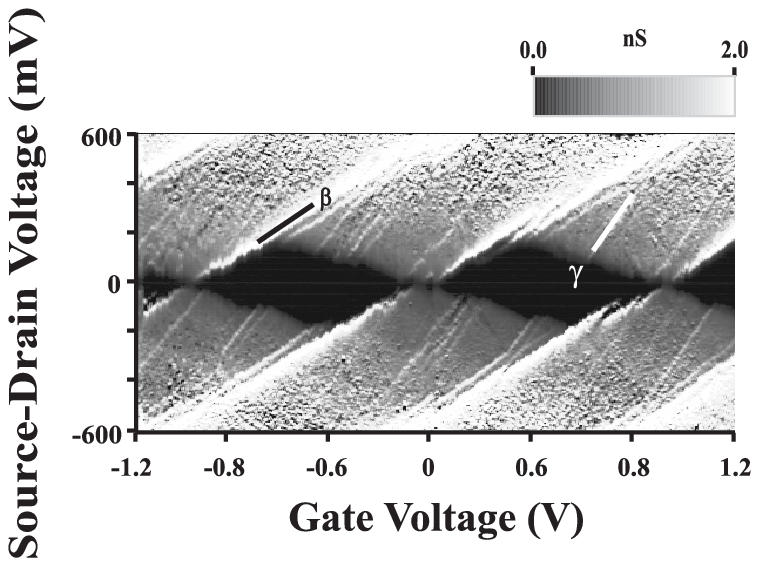}
\caption{SET with a charging energy $\sim 90$~meV. 
Non-orthodox lines $\gamma$ cannot be attributed to the recharging of some other grains.
}
\label{Exper}
\end{center} 
\end{figure}


\begin{thebibliography}{2}

\bibitem{Klein}
D.L. Klein, R. Roth, A.K.L. Lim, A.P. Alivisatos, P.L. McEuen,
Nature {\bf 389}, 699 (1997).

\bibitem{Persson}
S.H.M. Persson, L.G.M. Olofsson, L.K. Gunnarsson, {\it Appl. Phys. Lett.} {\bf 74}, 2546 (1999).

\bibitem{Soldatov}
E.S. Soldatov, V.V. Khanin, A.S. Trifonov, D.E. Presnov, S.A. Yakovenko, G.B. Khomutov,
G.P. Gubin, V.V. Kolesov JETP. Lett. {\bf 64}, 556 (1996).

\bibitem{Park}
H.Park, J.Park, A.K.L. Lim, E.K. Anderson, A.P. Alivisatos, P.L. McEuen,
Nature {\bf 407}, 57 (2000).

\bibitem{Ralph95}
D.C. Ralph, C.T. Black, M. Tinkham,
Phys. Rev. Lett. {\bf 74}, 3241 (1995).

\bibitem{Ralph97}
D.C. Ralph, C.T. Black, M. Tinkham,
Phys. Rev. Lett. {\bf 78}, 4087 (1997).

\bibitem{Gueron}
S. Gueron, M.M. Deshmukh, E.B. Myers, D.C. Ralph,
Phys. Rev. Lett. {\bf 83}, 4148 (1999).

\bibitem{Davidovich}
D. Davidovi\' c, M. Tinkham,
Phys. Rev. Lett. {\bf 83}, 1644 (1999).

\bibitem{OurAPL}
S.E. Kubatkin, A.V. Danilov, A.L. Bogdanov, H. Olin, T. Claeson,
Appl. Phys. Lett. {\bf 73}, 3604 (1998).

\bibitem{OurJLTP}
S.E. Kubatkin, A.V. Danilov, H. Olin, T. Claeson,
J. Low Temp. Phys. {\bf 118}, 307 (2000).

\bibitem{Ortho}
D.V. Averin and K.K. Likharev, in {\em Mesoscopic Phenomena in Solids\/},
(North-Holland, New York, 1991), Chap. 6.

\bibitem{Ruzin}
I.M. Ruzin, V. Chandrasekhar, E.I. Levin, L.I. Glasman,
Phys. Rev. B {\bf 45}, 13469 (1992).

\bibitem{Kouwenhoven}
L.P. Kouwenhoven, T.H. Oosterkamp, M.W.S. Danoesastro, M. Eto, D.G. Austing,
T. Honda, S. Tarucha, Science {\bf 278}, 1788 (1997).

\bibitem{Zeeman}
A conclusive argument is observation of the Zeeman shift for quantized energy levels in a magnetic 
field~\cite{Ralph97}, but to observe this effect in a magnetic field below 10~T one should 
lower the temperature to a dilution refrigerator range.

\bibitem{Schwabe}
N.F. Schwabe, A.N. Cleland, M.C. Cross, M.L. Roukes,
Phys. Rev. B {\bf 52}, 12911 (1995).

\bibitem{Gorelik}
L.Y. Gorelik, A. Isacsson, M.V. Voinova, B. Kasemo, R.I. Shekhter, M. Jonson,
Phys. Rev. Lett. {\bf 80}, 4526 (1998).

\bibitem{Simon}
URL: http://home1.gte.net/kittypaw/simon.htm.

\bibitem{Valles}
K.L. Ekinchi, J.M. Valles, Jr.,
Phys. Rev. B {\bf 58}, 7347 (1998);
K.L. Ekinchi, J.M. Valles, Jr.,
Phys. Rev. Lett. {\bf 82}, 1518 (1999).

\bibitem{Ruzin2}
A system of two grains connected in series was studied in~\cite{Ruzin}.
It was shown, that at temperatures below the smallest charging energy, 
the zero bias conductance, as a function of the gate voltage, is a set of randomly placed peaks with 
random amplitudes.


\end{thebibliography}
\end{document}